\documentclass[final,5p,times,twocolumn]{elsarticle}

\usepackage{amssymb}

\journal{Physica C}

\begin{document}

\begin{frontmatter}



\title{Pressure-Induced Antiferromagnetic Bulk Superconductor EuFe$_2$As$_2$}


\author[NIMS,TRIP]{Taichi Terashima}
\author[NIMS]{Hiroyuki S. Suzuki}
\author[NIMS]{Megumi Tomita}
\author[NIMS]{Motoi Kimata}
\author[NIMS]{Hidetaka Satsukawa}
\author[NIMS]{Atsushi Harada}
\author[NIMS]{Kaori Hazama}
\author[NIMS]{Takehiko Matsumoto}
\author[Osaka]{Keizo Murata}
\author[NIMS,TRIP]{Shinya Uji}

\address[NIMS]{National Institute for Materials Science, Tsukuba, Japan}
\address[Osaka]{Division of Molecular Materials Science, Graduate School of Science, Osaka City University, Osaka, Japan}
\address[TRIP]{JST, Transformative Research-Project on Iron Pnictides (TRIP), Tokyo, Japan}

\begin{abstract}
We present the magnetic and superconducting phase diagram of EuFe$_2$As$_2$ for $B \parallel c$ and $B \parallel ab$.  The antiferromagnetic phase of the Eu$^{2+}$ moments is completely enclosed in the superconducting phase.  The upper critical field vs. temperature curves exhibit strong concave curvatures, which can be explained by the Jaccarino-Peter compensation effect due to the antiferromagnetic exchange interaction between the Eu$^{2+}$ moments and conduction electrons.
\end{abstract}


\end{frontmatter}



Since the discovery of superconductivity (SC) at a transition temperature $T_c$ = 26 K in LaFeAsO$_{1-x}$F$_x$ by Kamihara \textit{et al}.\cite{Kamihara08JACS} extensive studies of SC in layered iron pnictides and related compounds have been performed.  There were some reports that the 122 parent compounds $A$Fe$_2$As$_2$ ($A$ = Ca, Sr, Ba, and Eu) could be tuned to SC by the application of high pressure $P$ \cite{Torikachvili08PRL, Park08JPCM, Alireza09JPCM, Fukazawa08JPSJ, Kotegawa09JPSJ, Igawa09JPSJ,  Lee08condmat, Miclea08condmat}.  However, most of these reports were based only on resistivity measurements and hence could not establish the bulk nature of $P$-induced SC.  Recently, we have proved from ac magnetic susceptibility measurements that $P$-induced \textit{bulk} SC does occur in EuFe$_2$As$_2$ and further that it coexists with the antiferromagnetic order of the Eu$^{2+}$ moments below $T_N$ ($< T_c$) \cite{Terashima09JPSJ_EFA}.  $P$-induced bulk SC has also been confirmed for SrFe$_2$As$_2$ \cite{Matsubayashi09JPSJ}.  The occurrence of the bulk SC in the parent compounds has an important implication: the key to SC in the iron pnictides is not carrier doping but suppression of the magnetic/structural transition.  Here we present $B_{appl}$ (applied magnetic field)-$T$ (temperature) phase diagrams of EuFe$_2$As$_2$ for $P \sim$ 26 kbar.

\begin{figure}[tb]
\begin{center}
\includegraphics[width=8cm]{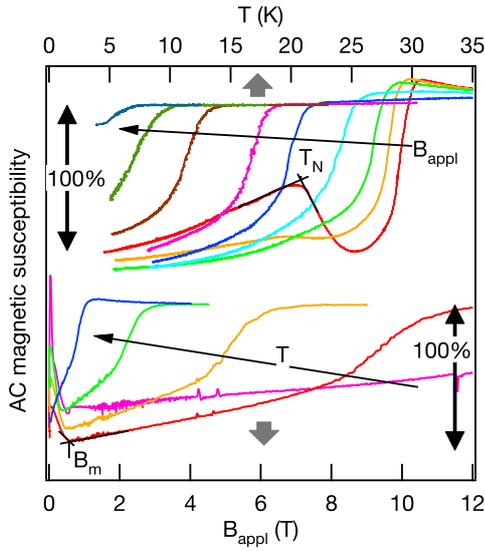}
\end{center}
\caption{(Color online) AC magnetic susceptibility along the $c$-axis of EuFe$_2$As$_2$ at $P$ = 25.7 kbar as a function of $T$ at a constant $B_{appl}$ (upper) and as a function of $B_{appl}$ at a constant $T$ (lower).   $B_{appl}$ was applied parallel to the $c$-axis.  The arrows indicate 100\% shielding.  For the upper curves, $B_{appl}$ = 0, 0.25, 0.5, 1, 2, 4, 8, 12, and 16 T.  For the lower curves, $T$ = 0.02, 10, 15, 20, and 25 K.  The definitions of $T_N$ and $B_m$ are shown.}
\label{chi_c}
\end{figure}

Figure~\ref{chi_c} shows ac magnetic susceptibility along the $c$ axis of EuFe$_2$As$_2$ at $P$ = 25.7 kbar.  (We recalibrated Manganin wire used for pressure gauges after the publication of Ref.~\cite{Terashima09JPSJ_EFA} and found that the pressure was estimated about 9\% larger in Ref.~\cite{Terashima09JPSJ_EFA}.  For the pressures of 28 and 29 kbar of the $c$-axis and $ab$-plane susceptibility measurements in Ref.~\cite{Terashima09JPSJ_EFA}, read 25.7 and 26.3 kbar, respectively.)  The susceptibility at zero field exhibits a large drop below $T_c$ = 31 K, then increases with decreasing $T$ to show a peak around $T_N$ = 21 K, where the Eu$^{2+}$ moments order antiferromagnetically, and finally decreases again.  The size of diamagnetism is consistent with 100\% shielding.  With increasing $B_{appl}$, the superconducting transition shifts to lower $T$.  The feature associated with $T_N$ is barely visible at $B_{appl}$ = 0.25 T, but is already absent at $B_{appl}$ = 0.5 T.  The susceptibility vs. $B_{appl}$ curves for $T < T_N$ show drops at low fields.  It indicates a field-induced transition from the antiferromagnetic to paramagnetic state of the Eu$^{2+}$ moments.  The characteristic field $B_m$ for this transition is defined in the figure.  (The definitions of $T_N$ and $B_m$ given in Fig.~\ref{chi_c} are different from conventional ones.  They have been adopted for practical purposes.)  In the case of the $a$-axis susceptibility, the zero-field curve shows a plateau below $T_N$, and hence we define $T_{N2}$ as the lower end temperature of the plateau.

 \begin{figure}[tb]
\begin{center}
\includegraphics[width=8cm]{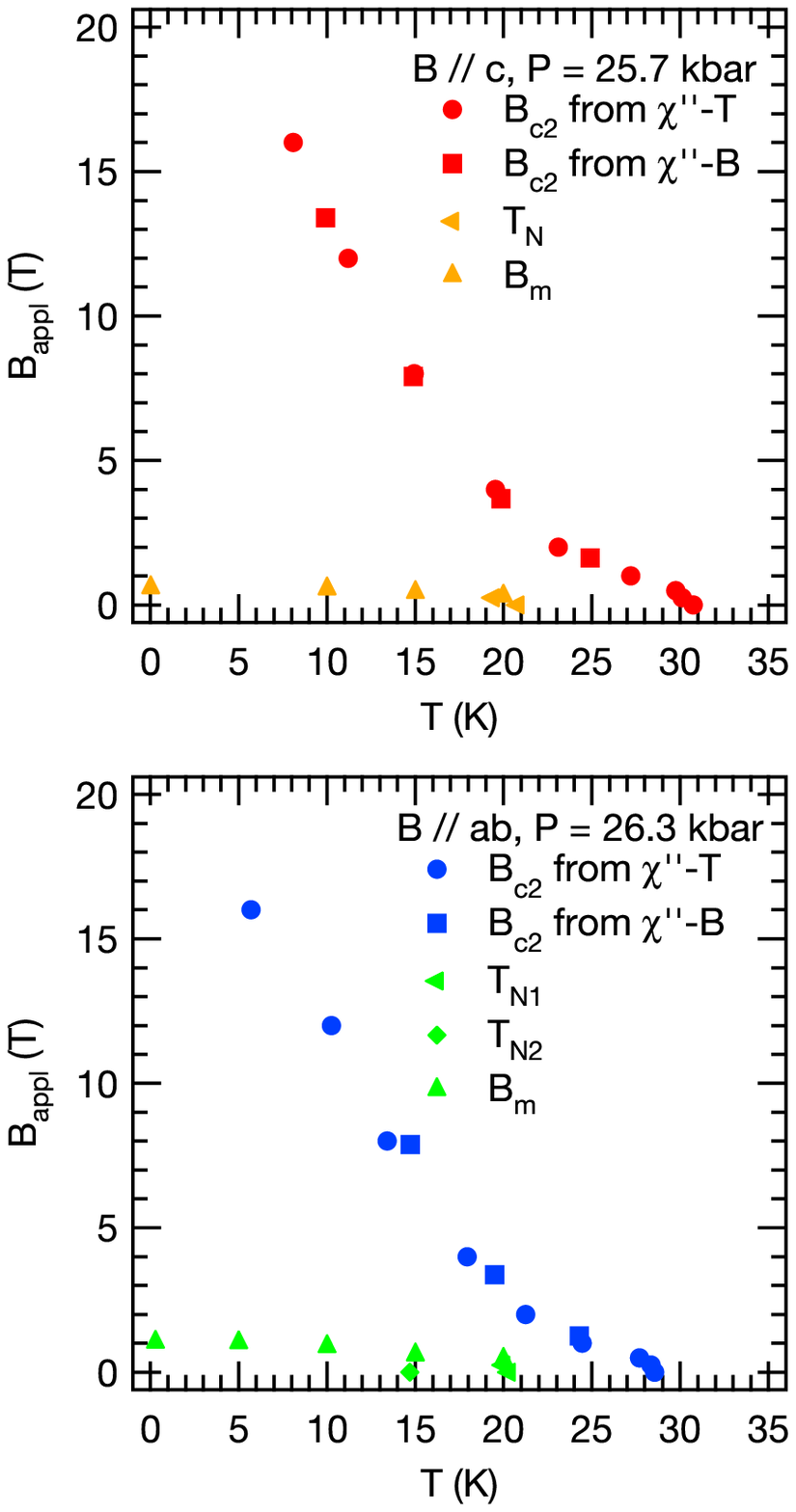}
\end{center}
\caption{(Color online) $B_{appl}$-$T$ phase diagrams of EuFe$_2$As$_2$.}
\label{phase}
\end{figure}

Figure~\ref{phase} shows the $B_{appl}$-$T$ phase diagrams for $B \parallel c$ and $B \parallel ab$.  The antiferromagnetic phase is completely enclosed within the superconducting phase.  The upper critical field ``$B_{c2}$'' shows peculiar temperature dependence (the quotation marks attached to $B_{c2}$ indicate that $B_{c2}$ values are based on the applied field, not on the internal field).  Initially, ``$B_{c2}$'' increases rather slowly compared to $P$ or doping-induced SC in other 122 compounds, and then it starts to increase rapidly approximately below zero-field $T_N$, resulting in pronounced concave curvatures of the ``$B_{c2}$'' -$T$ curves.  This can be explained by the Jaccarino-Peter compensation effect \cite{Jaccarino62PRL}.  For the sake of qualitative discussion, we follow the approximation used in Ref.~\cite{Decroux82Book}: $H_{c2} = H^*_{c2} - a[H_{c2}-|H_J|]^2$, where $H_{c2}$ is the upper critical field, $H^*_{c2}$ the orbital critical field, $H_J$ the antiferromagnetic exchange field acting on the conduction electrons, and $a$ determines the strength of the paramagnetic pairbreaking.  Although EuFe$_2$As$_2$ orders antiferromagnetically, the predominant interaction between the Eu$^{2+}$ moments is (inplane) ferromagnetic one \cite{Jiang09NJP}.  Thus the susceptibility rapidly increases as $T$ approaches $T_N$ from above.  Since $H_J$ is proportional to the magnetization of the Eu$^{2+}$ moments, it also increases rapidly.  This rapid increase in $H_J$ suppresses $H_{c2}$ relative to $H^*_{c2}$ in the temperature range between $T_c$ and approximately $T_N$.  This explains the initial slow increase in ``$B_{c2}$''.  On the other hand, below $\sim$$T_N$ the magnetization saturates at low fields and hence $H_J$ in the above equation is approximately constant.  Then, as $T$ is lowered, $H_{c2}$ increases, which \textit{reduces} the pair breaking (as long as $H_{c2} < |H_J|$), which in turn increases $H_{c2}$, and so on.  This leads to a rapid rise in $H_{c2}$ below $\sim$$T_N$ as is observed.

In summary, we have determined the $B_{appl}$-$T$ phase diagrams of EuFe$_2$As$_2$ for $B \parallel c$ and $B \parallel ab$.  The peculiar temperature dependence of ``$B_{c2}$'' can be explained by the Jaccarino-Peter compensation effect due to the antiferromagnetic exchange interaction between the Eu$^{2+}$ moments and conduction electrons.



\end{document}